# Task Scheduling for Heterogeneous Multicore Systems

Zhuo Chen, and Diana Marculescu, *Fellow, IEEE*

**Abstract**—In recent years, as the demand for low energy and high performance computing has steadily increased, heterogeneous computing has emerged as an important and promising solution. Because most workloads can typically run most efficiently on certain types of cores, mapping tasks on the best available resources can not only save energy but also deliver high performance. However, optimal task scheduling for performance and/or energy is yet to be solved for heterogeneous platforms. The work presented herein mathematically formulates the optimal heterogeneous system task scheduling as an optimization problem using queueing theory. We analytically solve for the common case of two processor types, e.g., CPU+GPU, and give an optimal policy (CAB). We design the GrIn heuristic to efficiently solve for near-optimal policy for any number of processor types (within **1.6%** of the optimal). Both policies work for any task size distribution and processing order, and are therefore, general and practical. We extensively simulate and validate the theory, and implement the proposed policy in a CPU-GPU real platform to show the optimal throughput and energy improvement. Comparing to classic policies like load-balancing, our results range from **1.08x~2.24x** better performance or **1.08x~2.26x** better energy efficiency in simulations, and **2.37x~9.07x** better performance in experiments.

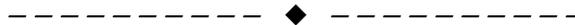

## 1 INTRODUCTION

With the ever-growing demand for high performance and low energy computing, heterogeneous multicore systems have emerged as a promising solution to meet this demand. In this work, we focus on single chip multicore heterogeneous systems, rather than datacenter-scale computing systems [42,43,44]. Prior work [8,9,10,17] has shown that specialized processors or accelerators, e.g., GPUs and FPGAs, can outperform traditional CPUs in terms of performance and energy for certain computational workloads. Therefore, combining different types of processing cores onto the same platform can better fit various application workloads and thus, will not only boost performance but also save energy. As a result, Open Computing Language (OpenCL) [18,19,20] has been developed to enhance the programmability of heterogeneous systems. Various heterogeneous system runtime environments and domain-specific languages that support heterogeneous systems [21,27,40,41] have been proposed as well. In OpenCL, Programmers are now able to program once and then run applications on heterogeneous platforms consisting of various types of processing units. However, task scheduling in heterogeneous multi-core systems is more difficult due to the inherent *affinity* of tasks to certain types of resources or processors. Finding out the (near-) optimal scheduling policy is important since it can be easily incorporated into existing heterogeneous system frameworks [18,19,20,21,27,40,41] and improve the overall system performance.

## 2 RELATED WORK AND PAPER CONTRIBUTIONS

There is a large amount of work on heterogeneous system performance and energy optimization [3,4,5,6,7]. Some approaches [6,7] optimize for the system performance statically. However, such approaches require extensive offline profiling and have to re-do the optimization for each new application and new input data. Therefore, such approaches cannot take into consideration dynamic workload variations. Other approaches, e.g., [1,37,38,39] have used queueing theory or linear programming optimization to optimize system throughput and power, but only work in *non-affinity* problems, in which the system consists of *iso-ISA processors* with different speeds or energy profiles, as opposed to *different types* of computing resources. Non-affinity example systems include ARM's big.LITTLE [22] or NVIDIA's Tegra [24] – in this case, tasks favor the fastest processor if no power budget is imposed. On the other hand, in a non-iso-ISA, true heterogeneous system (e.g., CPU-GPU), some tasks are inherently more suitable to run on CPU and others on GPU. Finally, prior work also addressed domain-specific heterogeneous system scheduling, e.g., proximity queries [34].

For affinity-based, true heterogeneous systems, there are also many theoretical results trying to solve for the *optimal* task scheduling policy by using queueing theory [13,14,15]. However, most of them only work with either Processor-Sharing (PS) or First-Come-First-Serve (FCFS) [13,29,30] processing order, and require Markovian assumptions, e.g., task arrivals modeled as Poisson processes and exponentially distributed task sizes, while these assumptions are often not satisfied, as correctly being pointed out before [11] [12]. In addition, they can only provide approximations via either computational [13] or analytical methods [2]. Ahn et al. [22] proposed a myopic policy that guarantees optimality under certain condi-

- *Zhuo Chen and Diana Marculescu are with Carnegie Mellon Univeristy, Pittsburgh, PA 15213.*
- *E-mail: zhuoc1@andrew.cmu.edu, dianam@cmu.edu.*

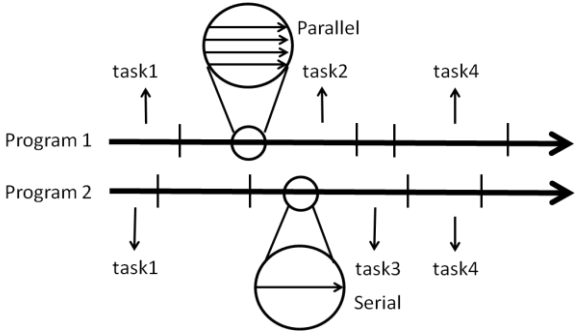

**Figure 1: We model the program as a single sequence of tasks, with each task exhibiting its own parallelism.**

tions by assuming no further arrivals. Bell et al. [28] demonstrate asymptotic optimality in heavy traffic regimes. Methods using learning algorithms [4,6,21] require that real world applications should have the same distribution as the training set (which is unrealistic), cannot guarantee optimality, and need huge online or offline training time. Accordingly, the approaches proposed so far mostly rely on restrictive assumptions, heuristics, approximations, machine learning techniques, and/or traditional scheduling policies to approximate the optimal task scheduling. None of them is able to prove policy optimality in a general case, even for a simple system with two different processor types.

In this work, we formulate the optimal task scheduling policy for heterogeneous systems with an arbitrary number of processor types as an *integer non-linear optimization* problem. We analytically determine the optimal policy, CAB (Choose-between-Accelerate-the-fastest-and-Best-fit), for heterogeneous systems with two types of processors. For the general case of arbitrary number of processor types, we design the GrIn (Greedy-Increase) heuristic that can solve efficiently for nearly optimal solutions. Extensive simulations and real platform experiments validate the optimality and generality of our proposed policy.

To the best of our knowledge, our work makes the following contributions:
1. We are the first to mathematically determine an *optimal scheduling policy* (*CAB*) for system performance and energy in a heterogeneous multi-core system with two types of processors. While some results follow the intuition, we also discover cases that give counter-intuitive results.
2. From the mathematical structure of the optimization problem, we design the GrIn (Greedy-Increase) algorithm that can find in *quadratic* time a nearly optimal solution which is within **1.6%** of the optimal solution for a heterogeneous multi-core system with *an arbitrary number* of processor types.
3. Our solution works under any task size distribution and processing order. No arrival process is involved here. Therefore, it is *very general* and, at the same time, *practical*. When compared to a generic solver, our approach is **up to 2x** faster, thereby indicating that a low overhead implementation is feasible.
4. We extensively *simulate and validate* our proposed policy and show that it indeed matches the theoretical results and outperforms all the other commonly used policies. Our results show that **1.08x** to **2.24x** better performance or **1.08x** to **2.26x** better energy efficiency can be achieved when compared to classic policies like load balancing.
5. We implement CAB, which is equivalent to GrIn algorithm under the case of two processor types, in a real CPU-GPU platform. The experimental results match the theoretical ones and provide the best performance, improving classic policies like load balancing by **2.37x** to **9.07x** in performance.

The rest of the paper is organized as follows. In section 3, we mathematically determine the optimal task scheduling policy CAB. In section 4, we extend it to include multiple types of processors and design the GrIn algorithm. Section 5 and 6 extensively simulate and validate the correctness and generality of CAB and GrIn, respectively. Section 7 gives the experimental results of a real CPU-GPU platform. Section 8 concludes this paper and discusses future work.

## 3 OPTIMAL HETEROGENEOUS TASK SCHEDULING FOR TWO PROCESSOR TYPES

### 3.1 Abstraction Level and Queueing Theory

Typical workloads exhibit different levels of parallelism, e.g., Instruction Level Parallelism (ILP) and Thread Level Parallelism (TLP). In this work, we model a program as a single sequence of tasks [35][36]. Each task can exhibit its own parallelism as shown in the example in Figure 1: two programs, each of which consists of a sequence of tasks with arbitrary execution times. Any particular task can be either parallel (e.g., task2 in program1), or sequential (e.g., task3 in program2). Tasks within the same program have to be executed sequentially due to the data dependencies. A real world example can be an OpenCL [25] program. It consists of a (usually sequential) host code running on the host processor and some (usually parallel) kernels being dispatched to other processors. The host code and the kernels are executed in a sequential order at the task level, like in Figure 1. We note that the task size (or execution time) can vary a lot across programs and depends on many factors, e.g., input data size, the processor it runs on, etc. This is one of the reasons that make such a system hard to model via queueing theory. Indeed, queueing theory usually assumes that task size follows a certain distribution, e.g., exponential or bounded Pareto [12][15], which more often than not, are not applicable.

In this work, we assume that, during a certain period of time, the total number of programs $N$ running on the heterogeneous platform is relatively stable. Such an assumption corresponds to a closed system, one in which the number of programs is fixed. This is not a restrictive assumption, as it can be relaxed to include piece-wise closed systems which is the case in practice, indeed reflecting the case of real life applications: on personal computers, applications are not launched and terminated very frequently. Similarly, in data centers, the tasks to be processed are statistically stable at different time scales [11,31]. If $N$ programs are running, there will always be $N$ tasks in the system. The programs are loaded in the memory before launching, therefore whenever one task is finished, the next task of that program will immediately be issued to a processor. Accordingly, at any time, the number of tasks in the system $N_{task}$ is the same as the number of programs $N$, and therefore, we use $N$ to denote both of them.

In summary, the modeling framework assumed is as follows: when $N$ programs are running, there will always be $N$ tasks being processed in the system. Whenever a task is finished, the next task is immediately sent into the system. This new task can be sent to one of the processors based on the task scheduling policy. This fits the model of a *closed batch network* as shown in Figure 2. In Figure 2, we assume $N$ programs, and two types of processors (or a cluster of processors thereof). We use *processor-sharing* (PS) (also known as time-sharing) as an example to do derivations in the sequel, although our results do not require any specific processing order for the processors.

Note that although this is a "closed" network, it does not require sending the completed task back to the system. The system is still "closed" as long as there is a task arriving in the system when a task is completed. This reflects the reality of how programs run in the computer system: a new task arrives when the old task is finished. We define the system throughput and task response time:

**Definition 1** (System throughput). System throughput $X$ is defined as the rate of task completion for the system: $X = \frac{Number\ of\ tasks\ completed}{Elapsed\ time}$

**Definition 2** (Task response time). The response time $T$ in a closed batch network is the time from entry to completion.

Little's Law [15] for a closed batch network connects together the system throughput and the task mean response time $\mathbb{E}[T]$. The law is very general and powerful because it makes no assumptions on the task arrival process, task size distribution, the network topology, or processing order. For a closed batch network, Little's Law states that:

$$N = X \cdot \mathbb{E}[T] \quad (1)$$

Since $N$ is a fixed number, $X$ and $\mathbb{E}[T]$ are inversely proportional to each other. Maximizing the system throughput $X$ is equivalent to minimizing the task mean response time $\mathbb{E}[T]$. Note that this is not the case in an open queueing network.

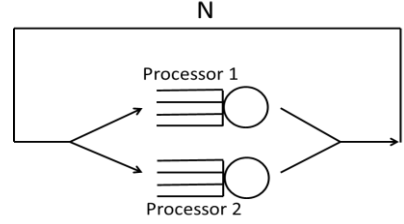

**Figure 2: A closed batch network. There are $N$ programs and two types of processors in the system. Whenever a task is completed, a new task will be sent in. We use processor-sharing (PS) processors as an example here.**

In the following sections, we will look for the optimal policy that maximizes the system throughput $X$. Per Little's Law, this automatically translates to minimizing the task mean response time.

### 3.2 Heterogeneous System Modeling

We first consider the case of a two-processor heterogeneous multi-core system as in Figure 2. This assumption will be relaxed later in the paper. There are two types of processors in the system: Processor1 (P1) and Processor2 (P2), each of which characterized by different Instruction Set Architectures (ISAs), which is the classic definition of heterogeneity in multi-core computing systems. For this heterogeneous-ISA system, a task will favor either one of the processors as pointed out in section 2. This property is called *task affinity* for a given processor type. Informally, if a task runs faster on P1, we call it a P1-type task, with affinity for P1. Conversely, if it is faster on P2, we call it a P2-type task, with affinity for P2. Because we have different processing rates in the heterogeneous system, we have the following definition:

**Definition 3** (Affinity matrix $\mu$). The affinity matrix is the task-processor matrix describing the processing rates of different types of tasks for each processor type: $\mu = \begin{matrix} & \begin{matrix} P1 & P2 \end{matrix} \\ \begin{matrix} \text{P1-type Task} \\ \text{P2-type Task} \end{matrix} & \begin{pmatrix} \mu_{11} & \mu_{12} \\ \mu_{21} & \mu_{22} \end{pmatrix} \end{matrix}$, where $\mu_{ij}$ represents the processing rate of i-type task on j-type processor.

By considering the task affinity, we always have:

$$\begin{cases} \mu_{11} > \mu_{12} \text{ (P1-type task is faster on P1 than on P2)} \\ \mu_{21} < \mu_{22} \text{ (P2-type task is faster on P2 than on P1)} \end{cases} \quad (2)$$

A concrete example can be P1 = CPU and P2 = GPU. Sequential tasks, e.g., sorting, usually have $\mu_{11} > \mu_{12}$ and therefore are CPU-type tasks, while tasks with high level of parallelism, e.g., image rendering, usually have $\mu_{21} < \mu_{22}$ and hence are GPU-type tasks.

Tasks can use resources differently on different processors, and therefore the power consumption of a task can also vary.

**Definition 4** (Power matrix $\mathcal{P}$). We use a power matrix to describe the processor power consumption when executing different types of tasks: $\mathcal{P} = \begin{matrix} & \begin{matrix} P1 & P2 \end{matrix} \\ \begin{matrix} \text{P1-type Task} \\ \text{P2-type Task} \end{matrix} & \begin{pmatrix} \mathcal{P}_{11} & \mathcal{P}_{12} \\ \mathcal{P}_{21} & \mathcal{P}_{22} \end{pmatrix} \end{matrix}$

So far, affinity has been defined in relation to performance only. To extend the concept to power consumption, from prior work [8][9] and considering the exponential power and performance relation from [26], we have

$\frac{\mathcal{P}_{iu}}{\mathcal{P}_{iv}} = \left(\frac{\mu_{iu}}{\mu_{iv}}\right)^\alpha$, $\alpha \leq 1$ for any type-i task, and $\alpha$ is a constant which is an average value for the system. This shows the relation between the performance ratio and its corresponding power ratio for a type-i task mapped onto processors u and v, respectively. $\alpha \leq 0$ indicates that the highest affinity processor for the task can provide both better performance and power (thus also better energy which is proportional to $\frac{1}{\mu} \cdot \mathcal{P}$); we call this a strong affinity regime. We call $0 < \alpha \leq 1$ a weak affinity regime, in which the highest affinity processor for the task can provide both better performance and energy, but worse power. Accordingly, we have $\mathcal{P}_{ij} = k\mu_{ij}^\alpha$ where k is a constant coefficient. The following two scenarios will be discussed in detail:

**Scenario 1** (Constant power). $\mathcal{P}_{ij} = $ constant k ($\alpha = 0$). This is the critical point between strong and weak affinity regimes.

**Scenario 2** (Proportional power). $\mathcal{P}_{ij} = k\mu_{ij}$ ($\alpha = 1$). In this case, power consumption is proportionally correlated with the affinity, e.g., a 100X faster device consumes 100X more power. This could result, e.g., from simply duplicating hardware resources.

We use system state matrix to describe the system task distribution: $\begin{smallmatrix}\text{P1-type Task}\\ \text{P2-type Task}\end{smallmatrix}\begin{pmatrix} \text{P1} & \text{P2} \\ N_{11} & N_{12} \\ N_{21} & N_{22} \end{pmatrix}$, where $N_{ij}$ is the number of i-type tasks within processor j's queue.

If we use $N_i$ to denote the total number of i-type tasks in the system, we have

$$\begin{cases} N_1 = N_{11} + N_{12} \\ N_2 = N_{22} + N_{21} \end{cases} \text{and } N = N_1 + N_2 \quad (3)$$

We assume that the type of each task is known before executing and hence $N_1$ and $N_2$ have known values. This can be done by characterizing or analyzing the tasks, and is only needed to be done once [4,7]. For example, a sequential task is usually CPU-type while highly parallel tasks are often GPU-type.

**Lemma 1.** *Given $N_1$ and $N_2$, the system state matrix only has two independent variables, e.g., $N_{11}$ and $N_{22}$, and the system throughput only depends on $N_{11}$ and $N_{22}$ as follows:*

$$X(N_{11}, N_{22}) = \frac{\mu_{11}}{N_{11} + N_2 - N_{22}}N_{11} + \frac{\mu_{21}}{N_{11} + N_2 - N_{22}}(N_2 - N_{22}) + \frac{\mu_{22}}{N_{22} + N_1 - N_{11}}N_{22} + \frac{\mu_{12}}{N_{22} + N_1 - N_{11}}(N_1 - N_{11}) \quad (4)$$

*Each $(N_{11}, N_{22})$ pair only corresponds to one system throughput value and one system task distribution (system state matrix).*

**Proof.** First, by using equation (3), we can see that the system state matrix only has two independent variables. As pointed out in section 2.1, we use PS processors as an example. The time-shared service rate $\mu_{ij}^*$ for each i-type task in each processor j is:

$$\mu_{ij}^* = \frac{\mu_{ij}}{\text{Number of tasks on processor } j} = \frac{\mu_{ij}}{\sum_{i=1}^{2} N_{ij}} \quad (5)$$

Processor j's throughput in terms of i-type tasks is:

$$X_{ij} = \mu_{ij}^* \cdot N_{ij} \quad (6)$$

Processor j's total throughput is:

$$X_j = \sum_{i=1}^{2} X_{ij} = \sum_{i=1}^{2} \mu_{ij}^* \cdot N_{ij} \quad (7)$$

Substituting (22) in (24), we obtain:

$$\begin{cases} X_1 = \frac{\mu_{11}}{N_{11} + N_{21}}N_{11} + \frac{\mu_{21}}{N_{11} + N_{21}}N_{21} \\ X_2 = \frac{\mu_{22}}{N_{22} + N_{12}}N_{22} + \frac{\mu_{12}}{N_{22} + N_{12}}N_{12} \end{cases} \quad (8)$$

Substituting (3) in (8), and summing up $X_1$ and $X_2$, we obtain (4).

Therefore, each $(N_{11}, N_{22})$ value determines one system throughput value and one system state matrix. □

Therefore, we can define the system state as follows:

**Definition 5** (System state $\mathcal{S}$). According to Lemma 1, we define the system state $\mathcal{S}$, which describes an unique system state, as $\mathcal{S} = (N_{11}, N_{22})$, where $N_{11} = 0, \ldots, N_1$ and $N_{22} = 0, \ldots, N_2$

### 3.3 Optimal Performance

For now we assume that the task size follows exponential distribution, which we will relax later. Using the Definition 5 of system state $\mathcal{S}$, we draw the Continuous-Time Markov Chain (CTMC) as in Figure 3.

The CTMC for the closed batch network consists of $N_\mathcal{S} = (N_1 + 1)(N_2 + 1)$ bubbles and each bubble represents a unique system state in $\mathcal{S}$ given by the number of tasks on the processors with the same type. The directed arcs denote possible transitions between any two adjacent states. The $\mu$ value on each arc is the processing rate of the tasks, and the $r$ value is the probability that the scheduler sends the next coming task to one of the processors.

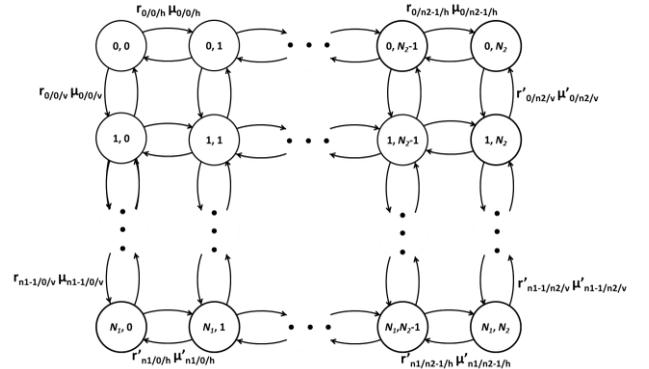

**Figure 3: CTMC of the closed batch network. Each bubble represents a system state $\mathcal{S}$. Here we temporarily assume exponentially distributed task size.**

Hence, $r$ determines the task scheduling policy. Both $\mu$ and $r$ are determined by the starting and ending states of the arc. For example, from state (0, 0) to state (0, 1), $\mu_{0/0/h}$ is the processing rate of P1 on P2-tasks and $r_{0/0/h}$ is the probability of sending P2-tasks to P2. It represents a transition from "all P2-tasks in P1 and all P1-tasks in P2" to "one P2-task in P2 and all P1 tasks in P2".

The general method of obtaining the r values is: (i) List all the balance equations of the CTMC in Figure 3; (ii)

Solve for the limiting probability $p(S)$ of each state $S$ in terms of the r values; (iii) Write down the system throughput as:

$$X_{sys} = \sum_{S=1}^{N_S} p(S) \cdot X(S) \quad (9)$$

where $X(S)$ is the throughput of a state $S$ and can be calculated by using Equation (4); and (iv) Find out the r values that maximize $X_{sys}$.

After some algebraic manipulations, one can obtain the r values, and thus determine the scheduling policy that maximizes the system throughput. However, we are able to prove the following lemma which helps us quickly obtain the optimal policy.

**Lemma 2.** *The maximum throughput is achieved when system always stays in the state $S_{max}$ that maximizes $X(S)$.*

**Proof.** For the system throughput $X_{sys}$ in (9), we notice that

$$X_{sys} = \sum_{S=1}^{N_S} p(S) \cdot X(S) \leq \sum_{S=1}^{N_S} p(S) \cdot X_{max} = X_{max} \quad (10)$$

where $X_{max} = \max(X(S))$ is the maximum throughput among all states and the corresponding $S_{max} = \text{argmax}_S X(S)$. This explains the inequality. The second equality comes from CTMC limiting probabilities adding up to one. Therfore, $X_{max}$ is achieved if we always stay in $S_{max}$. □

Previously, we assumed exponentially distributed task size. However, the above results hold for any task size distribution and processing order as stated in Lemma 3.

**Lemma 3.** *The maximum system throughput is achieved by staying in the state $S_{max}$, regardless of the task size distribution and processor's processing order.*

**Proof**. Without the exponential distribution assumption, we still have $N_S$ states and the state probabilities $p(S)$, though not the CTMC limiting probabilities. Equation (10) holds because $p(S)$ still sum to one. Therefore, again, the maximum system throughput is obtained by staying in the maximum throughput state $S_{max}$.

Here we consider the processing orders that are work-conserving, i.e., the processors keep working whenever there are tasks in the queue. Most commonly used processing orders, e.g., FCFS, PS, Last-Come-First-Serve (LCFS), etc., are work-conserving. Because the system always stays in the maximum throughput state, any work-conserving processing order will complete the same amount of work when time $\to \infty$, which means the time average throughput X will be the same for all of them. Accordingly, any processing order gives the same average throughput X. □

Therefore, to maximize the overall system throughput $X_{sys}$, we only need to find the state $S_{max} = \text{argmax}_S X(S)$, and we have the following lemma:

**Lemma 4.** *The optimal performance policy is achieved by assigning tasks to processors such that $S = S_{max}$ at any given time, where $S_{max}$ is determined by the element relations (not exact values) of the affinity matrix, as listed in Table 1.*

**Proof.** First, we assume $N_{11} \in [0, N_1]$ and $N_{22} \in [0, N_2]$ and take partial derivative of $X(S)$ with respect to $N_{11}$ and $N_{22}$:

$$\frac{\partial X}{\partial N_{11}} = \frac{(\mu_{11} - \mu_{21})(N_2 - N_{22})}{(N_{11} + N_2 - N_{22})^2} + \frac{(\mu_{22} - \mu_{12})N_{22}}{(N_{22} + N_1 - N_{11})^2} \quad (11)$$

$$\frac{\partial X}{\partial N_{22}} = \frac{(\mu_{11} - \mu_{21})N_{11}}{(N_{11} + N_2 - N_{22})^2} + \frac{(\mu_{22} - \mu_{12})(N_1 - N_{11})}{(N_{22} + N_1 - N_{11})^2} \quad (12)$$

If we let (11) and (12) equal zero, we have

$$\frac{(\mu_{11} - \mu_{21})(N_2 - N_{22})}{(N_{11} + N_2 - N_{22})^2} = -\frac{(\mu_{22} - \mu_{12})N_{22}}{(N_{22} + N_1 - N_{11})^2} \quad (13)$$

$$\frac{(\mu_{11} - \mu_{21})N_{11}}{(N_{11} + N_2 - N_{22})^2} = -\frac{(\mu_{22} - \mu_{12})(N_1 - N_{11})}{(N_{22} + N_1 - N_{11})^2} \quad (14)$$

If we divide (13) by (14), and also add (13) and (14) together, we can solve the equations to obtain:

$$\begin{cases} N_{11} = \dfrac{\mu_{11} - \mu_{21}}{\mu_{11} - \mu_{22} + \mu_{12} - \mu_{21}} N_1 \\ N_{22} = \dfrac{\mu_{12} - \mu_{22}}{\mu_{11} - \mu_{22} + \mu_{12} - \mu_{21}} N_2 \end{cases} \quad (15)$$

We obtain the following cases:

a. $\mu_{11} - \mu_{22} + \mu_{12} - \mu_{21} = 0$
  (1) $\mu_{11} = \mu_{22} = \mu_{12} = \mu_{21}$ (Homogeneous system)
    $X_{max} = X = \mu_{11} + \mu_{22}$, $S_{max}: -N_1 < N_{22} - N_{11} < N_2$
  (2) $\mu_{11} = \mu_{21}$, $\mu_{22} = \mu_{12}$, $\mu_{11} \neq \mu_{22}$ (big.LITTLE-like system)
    $X_{max} = X = \mu_{11} + \mu_{22}$, $S_{max}: -N_1 < N_{22} - N_{11} < N_2$
  (3) $\mu_{11} = \mu_{22} \triangleq \mu_1, \mu_{12} = \mu_{21} \triangleq \mu_2, \mu_1 > \mu_2$ (Symmetric heterogeneous system[2])
    $X_{max} = 2\mu_1, S_{max} = (N_1, N_2)$
b. $\mu_{11} - \mu_{22} + \mu_{12} - \mu_{21} \neq 0$

Considering the heterogeneous system affinity constraints (2), the following are the four relations between the elements of the affinity matrix.

| | (1) | | (2) | | (3) | | (4) | |
|---|---|---|---|---|---|---|---|---|
| $\mu_{11}$ | > | $\mu_{12}$ | $\mu_{11}$ > $\mu_{12}$ | | $\mu_{11}$ > $\mu_{12}$ | | $\mu_{11}$ > $\mu_{12}$ | |
| ∨ | | ∨ | ∧ | ∧ | ∨ | ∧ | ∧ | ∨ |
| $\mu_{21}$ | < | $\mu_{22}$ | $\mu_{21}$ < $\mu_{22}$ | | $\mu_{21}$ < $\mu_{22}$ | | $\mu_{21}$ < $\mu_{22}$ | |

By deriving the Hessian matrix [2], we find that the solution (15) in cases of (b.1) and (b.2) are saddle points. Therefore, the maximum throughput is on the boundaries ($N_{11} = 0$ or $N_1$, or $N_{22} = 0$ or $N_2$). For case (b.3), equations (13) and (14) do not have a solution, and throughput is a monotonic function. The maximum throughput in this case is also achieved on the boundaries. Case (b.4) is not valid, because $\mu_{11}$ has to be both larger than and smaller than $\mu_{21}$. Therefore, we obtain the maximum throughput in each case as follows:

b.1:
$$X_{max} = \frac{N_1 - 1}{N - 1}\mu_{12} + \frac{N_2}{N - 1}\mu_{22} + \mu_{11},$$
$$S_{max} = (1, N_2) \quad (16)$$

b.2:
$$X_{max} = \frac{N_2 - 1}{N - 1}\mu_{21} + \frac{N_1}{N - 1}\mu_{11} + \mu_{22},$$
$$S_{max} = (N_1, 1) \quad (17)$$

b.3:
$$X_{max} = \mu_{11} + \mu_{22},$$
$$S_{max} = (N_1, N_2) \quad (18)$$

b.4: Invalid. $\mu_{11}$ is both larger and smaller than $\mu_{21}$ □

**Table 1: Optimal performance policy always stays in the state $S_{max}$ determined by the task affinity.**

| | | | |
|---|---|---|---|
| Non-affinity | Homogeneous | $\mu_{11} = \mu_{22}$ $= \mu_{12} = \mu_{21}$ | $\forall N_{11}, N_{22}$ satisfy $-N_1 < N_{22} - N_{11} < N_2$ |
| | big.LITTLE-like | $\mu_{11} = \mu_{21}$, $\mu_{22} = \mu_{12}$, $\mu_{11} \neq \mu_{22}$ | |
| Affinity | Symmetric | $\mu_{11} = \mu_{22} \triangleq \mu_1$, $\mu_{12} = \mu_{21} \triangleq \mu_2$, $\mu_1 > \mu_2$ | $S_{max}$ $= (N_{11}, N_{22})$ $= (N_1, N_2)$ |
| | General-symmetric | $\mu_{11} > \mu_{12}$ $\vee \quad \wedge$ $\mu_{21} < \mu_{22}$ | |
| | (P1-)Biased | $\mu_{11} > \mu_{12}$ $\vee \quad \vee$ $\mu_{21} < \mu_{22}$ | $S_{max}$ $= (N_{11}, N_{22})$ $= (1, N_2)$ |
| | (P2-)Biased | $\mu_{11} > \mu_{12}$ $\wedge \quad \wedge$ $\mu_{21} < \mu_{22}$ | $S_{max}$ $= (N_{11}, N_{22})$ $= (N_1, 1)$ |

We can see that which $S_{max}$ should be chosen only depends on the ordering of affinity matrix elements, regardless of the $N_1, N_2$ values. This will be shown in the experimental results. The optimal policy covers traditional homogeneous and big.LITTLE-like systems, in which we can schedule any number of tasks on each processor as long as the queues are not empty, i.e., $-N_1 < N_{22} - N_{11} < N_2$. This is because these are non-affinity systems. They have only one type of tasks (big.LITTLE-like) or one type of processors (homogeneous).

We are interested in the affinity systems as shown in the last four rows in Table 1. In symmetric and general-symmetric cases, the optimal policy assigns all the P1-type tasks to P1 and P2-type tasks to P2, which follows the intuition of sending the tasks to their most suitable processors. We call it Best-Fit (BF) policy.

In P1- and P2-Biased cases, the optimal policy runs only a single program on the fastest processor and puts all other programs on the other processors. This result is unexpected and may seem counter-intuitive. We call this the Accelerate-the-Fastest (AF) policy.

In summary, our optimal policy is choosing between AF and BF based on the task affinity relations. Therefore, we call it CAB (Choose-between-AF-and-BF) policy. This approach also has the following advantages:

1. CAB is largely a static policy in the biased cases, which means the scheduler will keep running the tasks of the same program on the same processor even though the task types are different within a program. This leads to minimum memory transfer penalty in terms of both performance and energy, which is critical in heterogeneous system [3][27], compared to other dynamic policies.
2. CAB only needs the relations among the elements of affinity matrix, not exact values. Therefore, it is more error tolerant and easier to measure in real applications. CAB can also serve as a guide of how to partition the programs, e.g., one can partition the program such that CAB chooses AF to minimize memory transfer.

### 3.4 Optimal Energy and Energy-Delay Product

We use $\omega_{ij}$ to denote the average execution time in processor j of an i-type task and therefore, $\omega_{ij} = \frac{1}{\mu_{ij}}$. Also, we use $\rho_{ij}$ to denote the fraction of tasks that are i-type and have completed on processor j. Based on (5) and (6), we obtain $\rho_{ij} = \frac{X_{ij}}{X} = \frac{\mu_{ij}^* \cdot N_{ij}}{X}$.

The expected energy per task is calculated as:

$$\mathbb{E}[\mathfrak{E}] = \sum_{i,j} \rho_{ij} \times \mathcal{P}_{ij} \times \omega_{ij} = \sum_{i,j} \frac{\mu_{ij}^* \cdot N_{ij}}{X} \mathcal{P}_{ij} \frac{1}{\mu_{ij}}$$
$$= \frac{1}{X}\left(\frac{N_{11}}{N_{11}+N_{21}}\mathcal{P}_{11} + \frac{N_{21}}{N_{11}+N_{21}}\mathcal{P}_{21} \right. \quad (19)$$
$$\left. + \frac{N_{22}}{N_{22}+N_{12}}\mathcal{P}_{22} + \frac{N_{12}}{N_{22}+N_{12}}\mathcal{P}_{12}\right)$$

According to Little's Law, the delay per task is:

$$\mathbb{E}[T] = \frac{N}{X} \quad (20)$$

Therefore, the Energy-Delay Product (EDP) is:

$$\text{EDP} = \frac{\mathbb{E}[\mathfrak{E}]N}{X} \quad (21)$$

**Lemma 5.** *For Scenarios 1 and 2 (see Section 3.2), maximizing system throughput is equivalent to minimizing system energy and EDP.*

**Proof.** Using (19):
1. If $\mathcal{P}_{ij} = k$

$$\mathbb{E}[\mathfrak{E}] = \frac{2k}{X} \text{ and EDP} = \frac{2kN}{X^2} \quad (22)$$

2. If $\mathcal{P}_{ij} = k\mu_{ij}$

$$\mathbb{E}[\mathfrak{E}] = k \text{ and EDP} = \frac{kN}{X} \quad (23)$$

As both of them are inversely proportional to the system throughput $X$,

$$\text{Maximum } X \Rightarrow \text{Minimum } \mathbb{E}[\mathfrak{E}] \text{ and EDP} \quad \square$$

**Lemma 6.** *CAB is optimal in performance, energy and EDP, under Scenarios 1 and 2, i.e., CAB is the optimal policy $\mathfrak{P}$:*

$$\mathfrak{P} = \operatorname*{argmax}_{\mathfrak{P}} X = \operatorname*{argmin}_{\mathfrak{P}} \mathbb{E}[\mathfrak{E}] = \operatorname*{argmin}_{\mathfrak{P}} \text{EDP} \quad (24)$$

**Proof.** The optimal performance policy delivers the maximum system throughput, and therefore, it also minimizes energy and EDP when we have constant power or proportional power according to Lemma 5. $\square$

**Lemma 7.** *In the general power model, i.e., $\mathcal{P}_{ij} = k\mu_{ij}^\alpha, \alpha \leq 1$, energy and EDP are bounded by the constant power and proportional power scenarios (Scenarios 1 and 2). CAB is optimal in performance and EDP for any $\alpha \leq 1$ for compute-bound workloads, that is when throughput $X \to \infty$. CAB also delivers optimal energy in the strong affinity regime, i.e., $\alpha \leq 0$ when $X \to \infty$.*

**Proof.** Note that $\mathbb{E}[\mathfrak{E}(\alpha)]$ is now a function of $\alpha$ and we already have $\mathbb{E}[\mathfrak{E}(0)] = \frac{2k}{X}$ and $\mathbb{E}[\mathfrak{E}(1)] = k$.

Because power $\mathcal{P}_{ij}$ is always positive, we have:
$$\begin{cases} \mathbb{E}[\mathfrak{E}(\alpha)] \leq \mathbb{E}[\mathfrak{E}(0)], \alpha \leq 0 \\ \mathbb{E}[\mathfrak{E}(0)] \leq \mathbb{E}[\mathfrak{E}(\alpha)] \leq \mathbb{E}[\mathfrak{E}(1)], 0 \leq \alpha \leq 1 \end{cases}$$
$$\Rightarrow \begin{cases} \mathbb{E}[\mathfrak{E}(\alpha)] \leq \frac{2k}{X}, \alpha \leq 0 \\ \frac{2k}{X} \leq \mathbb{E}[\mathfrak{E}(\alpha)] \leq k \; 0 \leq \alpha \leq 1 \end{cases}$$

Similarly, for EDP:
$$\begin{cases} \text{EDP}(\alpha) \leq \text{EDP}(0), \alpha \leq 0 \\ \text{EDP}(0) \leq \text{EDP}(\alpha) \leq \text{EDP}(1), 0 \leq \alpha \leq 1 \end{cases}$$
$$\Rightarrow \begin{cases} \text{EDP}(\alpha) \leq \frac{2kN}{X^2}, \alpha \leq 0 \\ \frac{2kN}{X^2} \leq \text{EDP}(\alpha) \leq \frac{kN}{X}, 0 \leq \alpha \leq 1 \end{cases}$$

Therefore, when $X \to \infty$, $\text{EDP}(\alpha) \to 0$ for any $\alpha$ value and $\mathbb{E}[\mathfrak{E}(\alpha)] \to 0$ in strong affinity regime, i.e., $\alpha \leq 0$. □

## 4 THE CASE OF MULTIPLE PROCESSOR TYPES

### 4.1 Generalized Optimization Problem

The above section gives the proof of the optimal scheduling policy, CAB, in the case of two types of processors. In the following, we extend our solution to include multiple types of processors. "Multiple types" here has a broad meaning. It can refer to non-iso-ISA systems, e.g., CPU+GPU, and also iso-ISA systems, e.g., ARM-like big.LITTLE systems. We consider all these cases systems with different processor types. Moreover, within each processor type, there can be multiple identical processors. These identical processors are considered as a single "super" processor or cluster of processors of that type.

If we have k types of tasks and l types of processors, the time-shared service rate $\mu_{ij}^*$ in equation (5) can be generalized as:
$$\mu_{ij}^* = \frac{\mu_{ij}}{\text{Number of tasks on processor } j} = \frac{\mu_{ij}}{\sum_{i=1}^{k} N_{ij}} \quad (25)$$

Similarly, processor j's total throughput (equation (7)) becomes:
$$X_j = \sum_{i=1}^{k} X_{ij} = \sum_{i=1}^{k} \mu_{ij}^* \cdot N_{ij} \quad (26)$$

The total throughput of the system is:
$$X_{sys} = \sum_{j=1}^{l} X_j = \sum_{j=1}^{l} \sum_{i=1}^{k} X_{ij}$$
$$= \sum_{j=1}^{l} \sum_{i=1}^{k} \mu_{ij}^* \cdot N_{ij} = \sum_{j=1}^{l} \sum_{i=1}^{k} \frac{\mu_{ij} N_{ij}}{\sum_{i=1}^{k} N_{ij}} \quad (27)$$

In equation (27), $\mu_{ij}$'s are known values while $N_{ij}$'s are the variables we want to solve for. Accordingly, our method is now converted to an integer non-linear optimization problem with linear constraints as follows:

$$\text{Maximize} \quad X_{sys} = \sum_{j=1}^{l} \sum_{i=1}^{k} \frac{\mu_{ij} N_{ij}}{\sum_{i=1}^{k} N_{ij}} \quad (28)$$

$$s.t. \begin{cases} \sum_{j=1}^{l} N_{ij} = N_i, & i = 1, \dots, k \\ N_{ij} \in \mathbb{Z}_{\geq 0}, & i = 1, \dots, k; j = 1, \dots, l \end{cases} \quad (29)$$

By solving equations (28) and (29), we can obtain the optimal task scheduling policy. However, integer non-linear optimization problems cannot be solved exactly in sub-exponential time. Although it is possible to exhaustively search for the optimal solution for small scale problems (both small in the number of processor types and number of tasks), even medium size problems become intractable. Furthermore, if we want to solve the problem on the fly in a piece-wise fashion, e.g., solve equations (28) and (29) when the number of tasks changes, a fast algorithm is needed.

### 4.2 GrIn (Greedy-Increase) Algorithm

By looking at the mathematical structure of the objective function (28), we are able to design an efficient algorithm, called GrIn (Greedy-Increase), which is proven to increase the throughput until it hits a local maximum. Experimental results show that this local maximum is usually close to the global maximum obtained by exhaustive search. We also compare GrIn with Sequential Least SQuares Programming (SLSQP) [32], which solves relaxed (continuous value) non-linear optimization problems with linear constraints. Results show that GrIn not only delivers better solutions but also runs faster than SLSQP. GrIn is proven more scalable because the improvements get better when the number of processor types increases.

Let us first denote
$$w_{ij} = \frac{N_{ij}}{\sum_{i=1}^{k} N_{ij}} \quad (30)$$

Then we have
$$\sum_{i=1}^{k} w_{ij} = \sum_{i=1}^{k} \frac{N_{ij}}{\sum_{i=1}^{k} N_{ij}} = 1 \quad (31)$$

Accordingly, we can rewrite equation (28) as weighted sum of $\mu_{ij}$ values.
$$\text{Maximize} \quad X_{sys} = \sum_{j=1}^{l} \sum_{i=1}^{k} \mu_{ij} w_{ij} \quad (32)$$

In equation (32), the first sum is over the elements within each column and we know that, in each column, $w_{ij}$ values sum to one. The original k × l matrix is compressed to a 1 × l vector. The final sum is done over elements of this vector to get the final system throughput. Let us first consider the problem without constraints (29), in which case all columns become independent. We call the maximum $\mu_{ij}$ value in column j as max j-col μ and use $i_{max}(j)$ to denote the row index of max j-col μ. It's obvious that the maximum $X_{sys}$ value is achieved when, for each column j, we choose $w_{i_{max}j} = 1$ and set all the rest $w_{ij}$'s, $i \neq i_{max}$ to be 0. When we consider the constraints (29), the problem becomes more complicated because (i) numerators of $w_{ij}$'s need to sum to $N_i$'s for each row; and (ii) $w_{ij}$ values along each row may have different denominators and we cannot simply convert constraints to $\sum_{j=1}^{l} w_{ij} = \text{const.}$

Therefore, we designed a fast algorithm GrIn, whose complexity is O(k × l), to find out the near-optimal integer solutions. Algorithm 2 shows the pseudocode of GrIn algorithm. It starts with an initial matrix N_init generated by Algorithm 1. N_init follows the intuition of the max j-col μ concept, but already satisfies the constraints. GrIn then greedily increases $X_{sys}$ by moving the tasks among the processors as stated in Lemma 8.

**Lemma 8.** *When GrIn moves one task, it guarantees that system throughput $X_{sys}$ increases as well.*

**Proof.** Because different types of tasks are independent, without loss of generality, we consider moving type p tasks among all processors. This action corresponds to shuffling one row, row p here, of the N matrix. Let us consider the values of $X_j = \sum_{i=1}^{k} \frac{\mu_{ij}N_{ij}}{\sum_{i=1}^{k} N_{ij}}, j = 1, \dots, l$ in equation (28), which is a weighted sum along one column of the processing rate matrix. If we move one p type task into processor j, we calculate the difference of the $X_j$ values after $(X_j^+)$ and before $(X_j)$ the move: $X_{df+} = X_j^+ - X_j$.

After adding one p-type task to processor j:

$$X_j^+ = \sum_{i=1}^{p-1} \frac{\mu_{ij}N_{ij}}{\sum_{i=1}^{k} N_{ij} + 1} + \frac{\mu_{pj}(N_{pj}+1)}{\sum_{i=1}^{k} N_{ij} + 1} + \sum_{i=p+1}^{k} \frac{\mu_{ij}N_{ij}}{\sum_{i=1}^{k} N_{ij} + 1}$$

$$= \frac{\mu_{pj}}{\sum_{i=1}^{k} N_{ij} + 1} + \sum_{i=1}^{k} \frac{\mu_{ij}N_{ij}}{\sum_{i=1}^{k} N_{ij} + 1} \quad (33)$$

$$= \frac{\sum_{i=1}^{k} N_{ij}}{\sum_{i=1}^{k} N_{ij} + 1} X_j + \frac{\mu_{pj}}{\sum_{i=1}^{k} N_{ij} + 1}$$

We then have:

$$X_{df+} = X_j^+ - X_j = \frac{\mu_{pj} - X_j}{\sum_{i=1}^{k} N_{ij} + 1} \quad (34)$$

Similarly, if we remove a p-type task from processor j:

$$X_j^- = \sum_{i=1}^{p-1} \frac{\mu_{ij}N_{ij}}{\sum_{i=1}^{k} N_{ij} - 1} + \frac{\mu_{pj}(N_{pj}-1)}{\sum_{i=1}^{k} N_{ij} - 1} + \sum_{i=p+1}^{k} \frac{\mu_{ij}N_{ij}}{\sum_{i=1}^{k} N_{ij} - 1}$$

$$= -\frac{\mu_{pj}}{\sum_{i=1}^{k} N_{ij} - 1} + \sum_{i=1}^{k} \frac{\mu_{ij}N_{ij}}{\sum_{i=1}^{k} N_{ij} - 1} \quad (35)$$

$$= \frac{\sum_{i=1}^{k} N_{ij}}{\sum_{i=1}^{k} N_{ij} - 1} X_j - \frac{\mu_{pj}}{\sum_{i=1}^{k} N_{ij} - 1}$$

We then have:

$$X_{df-} = X_j^- - X_j = \frac{X_j - \mu_{pj}}{\sum_{i=1}^{k} N_{ij} - 1} \quad (36)$$

If processor j has the highest $X_{df+}$ value, it indicates that a task move to processor j will increase the throughput most. If processor j has the lowest $X_{df-}$ value, it means that a task move from processor j provides the least throughput degradation.

Accordingly, if we only move tasks from $\max(X_{df-})$ processor to $\min(X_{df+})$ processor when $\max(X_{df+}) - \min(X_{df-}) \geq 0$, we are always increasing the system throughput $X_{sys}$. □

In Algorithm 2, GrIn first initializes N and then, for each row, i.e., each task type, it calculates the $X_{df+}$ and $X_{df-}$ values for all the processors (Line 5). This can be done in $O(k \times l)$ for all rows. We can then find the maximum value of $X_{df+}$ and minimum value of $X_{df-}$ (Line 6), which can be done in $O(k \times l)$ as well. Finally, we move one task from processor $\min(X_{df-})$ to processor $\max(X_{df+})$. Therefore, the complexity of GrIn is $O(k \times l)$.

Algorithm 1 gives the pseudocode of initializing the task distribution matrix N. To calculate the initial matrix, first we find out the max j-col μ values for each column. We then accordingly create a 0-1 max μ matrix $\mathfrak{U}$ of the same size $k \times l$, where the entries of those max j-col μ values are 1 in $\mathfrak{U}$ and remaining entries are 0s. There is only one 1 in each column of $\mathfrak{U}$, however, there may be multiple or zero 1s in each row of $\mathfrak{U}$.

The initializing algorithm works as follows: If there is only one 1 in the row of $\mathfrak{U}$ at position (i, j), we simply assign all the i-type tasks to processor j, which means $N_{ij} = N_i$ (Line 14-16). If there is more than one 1 value in row i, we sort the max j-col μ values accordingly, e.g., $\mu_{ij_1} > \mu_{ij_2} > \cdots > \mu_{ij_m}$. Starting from the largest μ value in that row, i.e., $\mu_{ij_1}$, we assign one task to processor $j_1$. The algorithm does the same for the remaining μ values in that row, except the last one $\mu_{ij_m}$. We assign all the remaining tasks, if any, to processor $j_m$ (Line 6-13). The last case corresponds to no 1s in row i. In this case, we first assign all tasks to processor i, and then use Lemma 8 to re-distribute the tasks until we reach a local maximum of system throughput (Line 18-21).

**Algorithm 1:** Pseudocode of initializing matrix N

1: **Input:** $k, l, N_i, \mu$
2: **Output:** $N\_init$
3: $N\_init \leftarrow$ diagonal matrix with value $N_i$
4: $\mathfrak{U}_{ind}$ = row index of maximum $\mu$ in each column
5: Create $k \times l$ matrix $\mathfrak{U}$ with 1s in $\mathfrak{U}_{ind}$ of each column and 0s for the rest.
5: **for** row = 1: $k$ **do**
6:     **If** number of 1s in row > 1 **do**
7:         $j_1, j_2, \dots, j_m$ = descendingly sort them based on their corresponding $\mu$ values
8:         leftTask = $N_i$[row]
9:         **for** mId = 1:m **do**
10:           $N[row, j_{mId}] = 1$; leftTask decreases by 1;
11:           break if leftTask == 0
12:         **end for**
13:         $N[row, j_m] = N[row, j_m] + leftTask$
14:     **elseif** number of 1s in row == 1 **do**
15:         Find j that $\mathfrak{U}[row, j] == 1$
16:         $N[row, j]$ = leftTask
17:     **else do**
18:         Calculate $X_{df+}$ and $X_{df-}$ of all processors [Eq.19, 21]
19:         Find $\max(X_{df+})$ and $\min(X_{df-})$
20:         $N[row, \min(X_{df-})]$ decreases by 1
21:         $N[row, \max(X_{df+})]$ increases by 1
22:     **end if**
23: **end for**
24: **return** $N\_init$

**Algorithm 2:** Pseudocode of GrIn

1: **Input:** $k, l, N_i, \mu$
2: **Output:** N
3: $N \leftarrow$ initial_N($k, l, N_i, \mu$)
4: **for** row = 1: $k$ **do**
5:     Calculate $X_{df+}$ and $X_{df-}$ of all processors [Eq.19, 21]
6:     Find $\max(X_{df+})$ and $\min(X_{df-})$
7:     $N[row, \min(X_{df-})]$ decreases by 1
8:     $N[row, \max(X_{df+})]$ increases by 1
9: **end for**

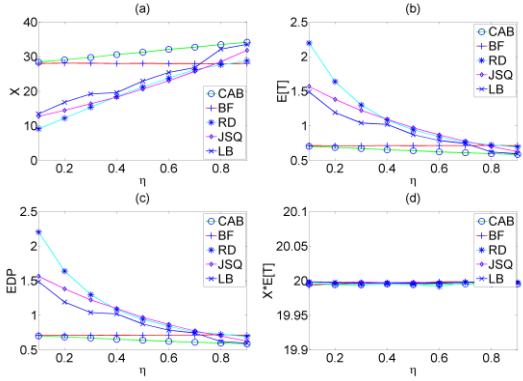

**Figure 4:** Four simulated metrics of all five policies under *exponentially* distributed task size. CAB gives the highest throughput and lowest mean response time and EDP.

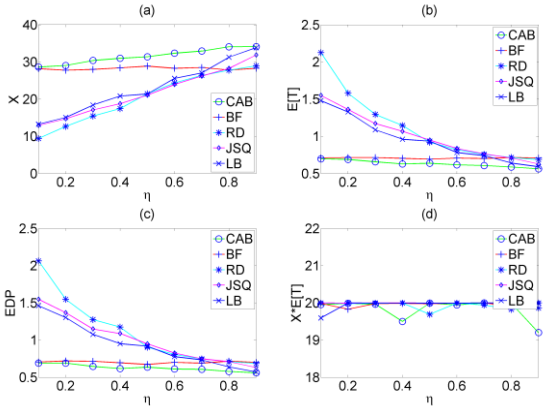

**Figure 5:** Four simulated metrics of all five policies under *bounded Pareto* distributed task size. CAB gives the highest throughput and lowest mean response time and EDP.

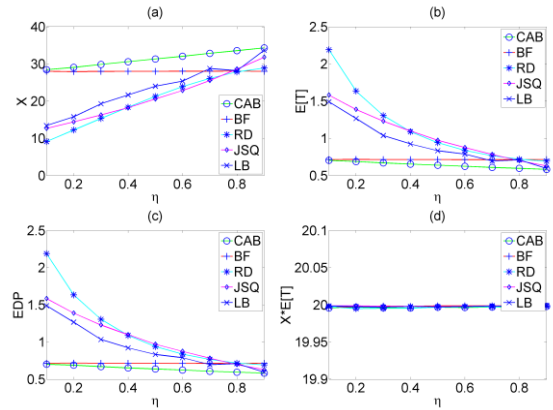

**Figure 6:** Four simulated metrics of all five policies under *uniformly* distributed task size. CAB gives the highest throughput and lowest mean response time and EDP.

   10:    **return** $N$

In the following sections, we extensively validate CAB and GrIn by using simulations and real CPU-GPU platform. We focus on affinity systems that illustrate the most interesting and representative cases. General-symmetric, P1- and P2-biased cases are all explored for two types of processors. GrIn algorithm is simulated against existing optimization method SLSQP and exhaustive search under systems with different number of processor types. We use PS in simulation and FCFS on a real platform to demonstrate the independence of the processing order. Different task size distributions are used to show the generality of our policy.

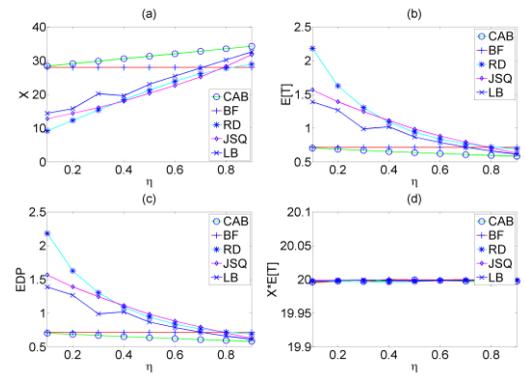

**Figure 7:** Four simulated metrics of all five policies under *constantly* distributed task size. CAB gives the highest throughput and lowest mean response time and EDP.

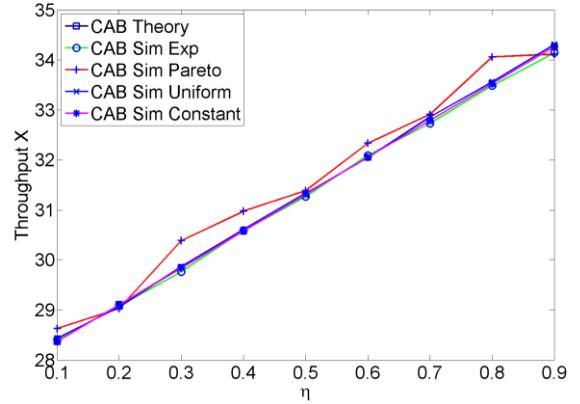

**Figure 8:** Comparison of the theoretical and simulated throughputs of CAB under four distributions. They are almost identical.

## 5. SIMULATION RESULTS FOR TWO TYPES OF PROCESSORS

In this section, we validate our theoretical results by extensively simulating a two-processor, e.g., CPU-GPU, heterogeneous system in an in-house heterogeneous system simulator. Task affinities are represented by the $\mu$ matrices. We launch $N = 20$ programs, and hence there are always 20 tasks running in the system. We use $\eta$ to denote the fraction of tasks that are P1-type, and thus $(1 - \eta)$ of the tasks belong to P2-type. We simulate over nine $\eta$ values from 0.1 to 0.9. Due to space limitations, we consider the proportional power model and show the most interesting and counter-intuitive P1-biased case with $\mu = \begin{pmatrix} 20 & 15 \\ 3 & 8 \end{pmatrix}$. Other cases are explored in the real platform experiments.

To demonstrate that our policy is independent on the task size distribution, we generate tasks by using the following distributions:

1. Exponential distribution, the Markovian property usually assumed in queueing theory
2. Bounded Pareto distribution, which many tasks follow [12,16]
3. Uniform distribution

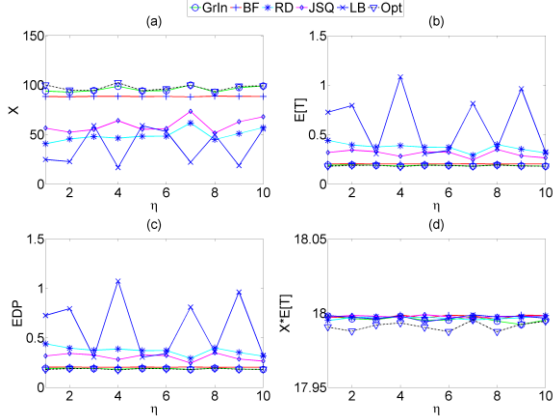

**Figure 9: Four simulated metrics of all six policies under *exponentially* distributed task size. GrIn gives higher throughput and lower mean response time and EDP than BF, RD, JSQ and LB. GrIn is very close to the optimal solution Opt obtained by exhaustive search.**

4. Constant distribution

We keep track of the number of tasks completed in a certain period of time to determine the simulated system throughput $X_{sim} = \frac{\text{Number of completed tasks}}{\text{Elapsed time}}$. We also record the response time T of each completed task and $\mathbb{E}[T_{sim}] = \frac{\Sigma_{i=\text{completed tasks}} T_i}{\text{Number of completed tasks}}$. Energy is the product of power ($\mathcal{P}$) and execution time ($\omega$), so we have $\mathbb{E}[\mathfrak{E}_{sim}] = \frac{\Sigma_{i=\text{completed tasks}} \mathcal{P}_i \cdot \omega_i}{\text{Number of completed tasks}}$ and $EDP_{sim} = \mathbb{E}[\mathfrak{E}_{sim}] \cdot \mathbb{E}[T_{sim}]$. Note that execution time of a task is not its response time, since response time includes the time waiting in the queue.

Because it is impossible to demonstrate optimality by simulating all possible policies (which are infinitely many) we compare CAB with the following commonly used policies:

1. Random (RD): randomly dispatch the task to the two types of processors with equal probability.
2. Best Fit (BF): dispatch the task to its "favorite" processor based on affinity.
3. Load Balancing with perfect information (LB): dispatch the task to balance the load of the processors, i.e., send it to the queue with the least amount of work. Work is defined as the task total size in the queue and is often estimated by assuming certain task size distribution. In our experiments, we use true task sizes which will only give better results than using estimations.
4. Join-the-Shortest-Queue (JSQ): dispatch the task to the processor with the least number of tasks.

Figures 4, 5, 6 and 7 show $X_{sim}$, $\mathbb{E}[T_{sim}]$, $EDP_{sim}$ and $X_{sim} \cdot \mathbb{E}[T_{sim}]$ for five policies across nine $\eta$ values, under *Exponential*, *Bounded Pareto*, *Uniform* and *Constant* distributions, respectively. According to Little's Law, $X_{sim} \cdot \mathbb{E}[T_{sim}]$ should always be $N = 20$ under any policy. The bottom right subplots verify this. Furthermore, as given in (23), $\mathbb{E}[\mathfrak{E}_{sim}]$ should be one and $EDP_{sim}$ should be the same as $\mathbb{E}[T_{sim}]$, in the proportional power case. This is demonstrated in bottom left subplots.

Figure 8 compares the theoretical throughput of CAB and the simulated CAB throughput under all four task size distributions. We see that the simulation results match well with the theoretical one. The reason of small deviation of bounded Pareto distribution will be discussed in the following.

From the figures above, we can see that:

1. CAB indeed always delivers the highest throughput and lowest mean response time and EDP among all the policies. Comparing to classic policies like load balancing, CAB delivers 1.08x to 2.24x better performance or 1.08x to 2.26x better energy efficiency (EDP). The exact improvement numbers vary with different µ matrices and $N_i$.
2. CAB is indeed independent of the task size distribution and processing orders for any η values.
3. The simulated results are almost the same as the theoretical results. The small error is due to the stochastic nature of the simulation, and the fact that we cannot run simulation for infinitely long time, which is assumed in queueing theory.
4. CAB and BF get very close when $\eta = 0.1$, because their states are quite similar: $S_{CAB} = (1,18)$, $S_{BF} = (2,18)$, and $X_{CAB} - X_{BF} = \frac{N_1-1}{N-1}(\mu_{12} - \mu_{22}) = \frac{\eta N-1}{N-1}(\mu_{12} - \mu_{22}) = 0.37$ which is, relatively, a very small difference.

Bounded Pareto is a heavy-tailed distribution. Therefore, its simulated results have a higher variance as shown in Figure 5 and 8. However, the variation can be reduced by simulating a longer time such that the heavy tail is sampled enough amount of times [16]. This is observed in our simulation.

## 6. SIMULATION RESULTS FOR MULTIPLE TYPES OF PROCESSORS

In this section, we show the effectiveness of GrIn by (i) simulating multiple types of processors with various policies under different task size distributions, and (ii) comparing GrIn with SLSQP and exhaustive search.

The existing optimization method we used is Sequential Least SQuares Programming (SLSQP) [32]. SciPy library of Python has an efficient implementation of SLSQP and it can solve for non-linear optimization problems with bounds, equality and inequality constraints such as equation (14).

In this section, we also simulate four commonly used policies, i.e., BF, RD, JSQ and LB, with four different task size distributions, i.e., exponential, bounded Pareto, uniform and constant. $X_{sim}$, $\mathbb{E}[T_{sim}]$, $EDP_{sim}$ and $X_{sim} \cdot \mathbb{E}[T_{sim}]$ metrics are reported for all the cases. Furthermore, the optimal policy (Opt) solved by exhaustive search is also shown in the figures for comparison.

Figure 9, 10, 11 and 12 show the simulation results for all polices under all four task size distributions. The x-axis denotes different sample points. The size of µ matrices is $3 \times 3$ since larger size takes significant time in exhaustive search. We randomize the entries of μ matrices and $N_i$ values to show the generality of GrIn for widely varying task affinities [8,9,10,17]. We performed 1,000 simulations and show 10 random samples of a random μ matrix (different $N_i$ values) here to maintain readability of the figures. For 1,000 runs, GrIn is constantly better than the other approaches and very close to the optimal solution. On average, GrIn is 1.6% from the optimal solution.

The results show that:
1. GrIn is indeed always better than the other commonly used policies in terms of throughput, response time and energy-delay product.
2. GrIn is independent of the task size distribution and processing order.
3. The system throughput determined by GrIn is only 1.6% from the optimal solution by exhaustive search, after averaging over 1,000 runs.
4. High variation of bounded Pareto distribution is again observed in Figure 10 and can be mitigated

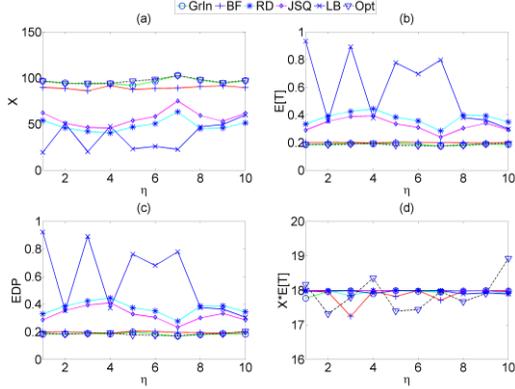

Figure 10: Four simulated metrics of all six policies under *bounded Pareto* distributed task size. GrIn gives higher throughput and lower mean response time and EDP than BF, RD, JSQ and LB. GrIn is very close to the optimal solution Opt obtained by exhaustive search.

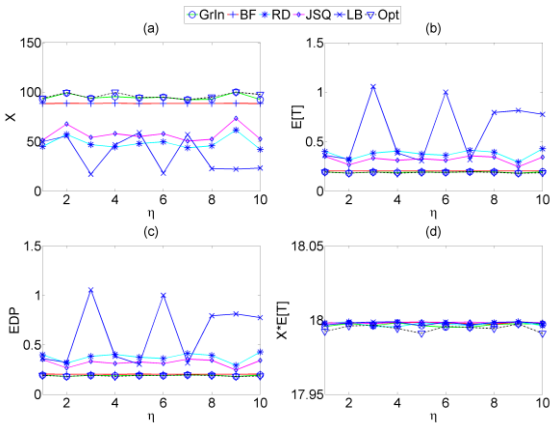

Figure 11: Four simulated metrics of all six policies under *uniformly* distributed task size. GrIn gives higher throughput and lower mean response time and EDP than BF, RD, JSQ and LB. GrIn is very close to the optimal solution Opt obtained by exhaustive search.

by simulating a longer time.

Figure 13 shows the comparison of solutions obtained by GrIn and SLSQP. The *y*-axis is the improvement of

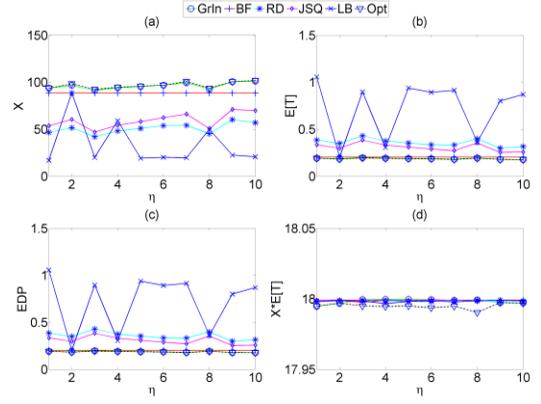

Figure 12: Four simulated metrics of all six policies under *constantly* distributed task size. GrIn gives higher throughput and lower mean response time and EDP than BF, RD, JSQ and LB. GrIn is very close to the optimal solution Opt obtained by exhaustive search.

GrIn over SLSQP. Since SLSQP solves for a relaxed optimization problem while GrIn solves for an integer optimization problem, SLSQP has a larger solution space to explore and can theoretically give better results than GrIn. We did not round the solution of SLSQP to integer values, because considering the constraints and high dimension of the objective function, converting to the integer solution is not a trivial task. We simulated across different μ matrix sizes, i.e., number of processor types, from $3 \times 3$ to $10 \times 10$, and for each size, we randomize the element values of μ matrices. The results are averaged over 100 runs under each matrix size, and they show that GrIn's integer solution is better than the continuous value solution of SLSQP. Moreover, GrIn performs better when we have more processor types. For example, GrIn achieves **5.7%** improvement over SLSQP in a system with ten different processor types.

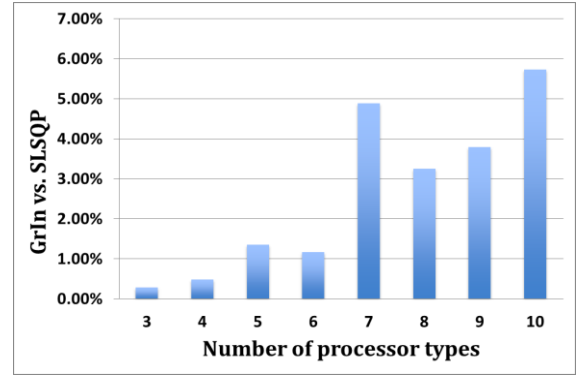

Figure 13: Improvement of GrIn's integer solution over SLSQP's continuous value solution. When we increase the number of processor types, GrIn delivers better results.

SLSQP has a drawback of being too restrictive on the objective function. It requires the objective function to be second-order differentiable. In our problem, the objective function equation (13) is discontinuous on the boundaries where the sum of any column of *N* matrix is zero. In the simulations, we do see SLSQP convergence failures.

Figure 14 shows the algorithm runtime comparison between GrIn and SLSQP. Since Python has an efficient SLSQP implementation, we also implement GrIn in Python. We note that:

1. Either one of them can report a smaller runtime by sacrificing the solution quality, e.g., one can use the initial guess as the solution, which can have bad quality, and thus report a very small runtime.
2. SLSQP sometimes fails to converge.

Because of these two observations, we only record the runs for which both approaches deliver similar throughput values (within 5% here). This approach gives a more reliable runtime for both algorithms when they can deliver similar solutions. For each number of processor types, 100 runs are simulated with randomized µ matrices and we report the average number as the final result.

From the results, we can see that GrIn is not only up to 2x faster than SLSQP but also more scalable as we increase the number of processor types. GrIn is almost twice as fast

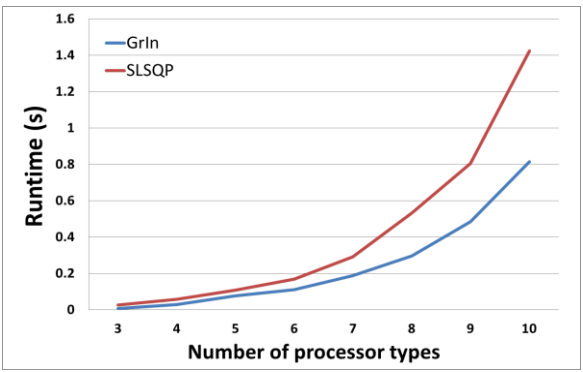

**Figure 14: Algorithm runtime comparison between GrIn and SLSQP. GrIn is not only faster but also more scalable.**

as SLSQP in a system with ten types of processors.

## 7. REAL PLATFORM RESULTS

To further demonstrate the efficacy of our policy, we implement it in OpenCL, a programming framework for heterogeneous systems, on a CPU-GPU real platform and measure its throughput.

Because (i) we only have machine with two types of processors and (ii) the GrIn algorithm gives the same solution as CAB's analytical solution in systems with two processor types, we do real-world experiments with two types of processors, i.e., CPU+GPU, under CAB policy. Different from simulation, here we use FCFS processing order to demonstrate CAB's generality. For simplicity, we used two benchmarks: (i) quick sort, which is sequential and thus favors CPU; and (ii) single layer Neural Network (NN), which has high parallelism and favors GPU. Although we are using two benchmarks here, the following method is generally applicable to any programs.

### 7.1 System Configuration

**Table 2: Real platform configuration**

| CPU | Intel Core i7-4790 Quad core, 8MB cache |
|---|---|
| GPU | NVIDIA GeForce GTX 760Ti 2GB DDR5 |
| System DRAM | 16GB |
| OS | Ubuntu 14.04.2 |

Table 2 shows the configuration of the CPU-GPU platform. Since the NVIDIA GPU has a 5-second auto timeout, we turn off the *lightdm* of Ubuntu to disable the timeout such that our program can run as long as we want to obtain statistically correct results. The algorithms are implemented in OpenCL, and we create context for each device, i.e., context1 for CPU and context2 for GPU. Each context has one single queue to implement the FCFS processing order. The kernels are generated from the two benchmarks mentioned above. Due to the limited access to the power sensors of CPU and GPU, we only report performance values here.

### 7.2 Processing Rate Measurement

First, we need to measure the average processing rate of each kernel on each processor. We run each kernel 1000 times and calculate the average execution time $\omega$, and therefore, the processing rate $\mu = 1/\omega$. One thing to note is that the input data size of the kernel can affect its running time and hence also the processing rate, e.g., it takes longer to sort 1,000 numbers than 500 numbers. The measured processing rates of the kernels with different input data sizes are listed in Table 3. We used these input data sizes since they are able to cover the most interesting and important cases: general-symmetric and P2-biased.

**Table 3: Measured processing rates of the kernels on both processors**

| Benchmark | Input size | $\mu_{CPU}$ ($s^{-1}$) | $\mu_{GPU}$ ($s^{-1}$) |
|---|---|---|---|
| quicksort-500 | 1,000 | 928 | 3.61 |
| quicksort-1000 | 3,500 | 253 | 0.911 |
| NN-2000 | 2000 | 587 | 2398 |

As discussed in Section 3.3, we do not need accurate values of processing rates here. It is sufficient to know their relative ordering.

### 7.3 P2-biased

According to the theoretical results, when the system is P2-biased, CAB will choose AF policy. Therefore, we pick quicksort-1000 and NN-2000. We launch N = 20 benchmarks and sweep over the nine η values.

Figure 15 shows the experimental throughput under different policies as well as the theoretical optimal (CAB) throughput of the system. Here CAB=AF is indeed optimal and is close to the theoretical value. AF (CAB) and BF are becoming closer when $\eta$ increases. This is because $X_{AF} - X_{BF} = \frac{N_2 - 1}{N - 1}(\mu_{21} - \mu_{11})$, $N_2$ decreases when $\eta$ increases. CAB outperforms the others, e.g., LB by 3.27x to 9.07x, and the experimental results match the theoretical results.

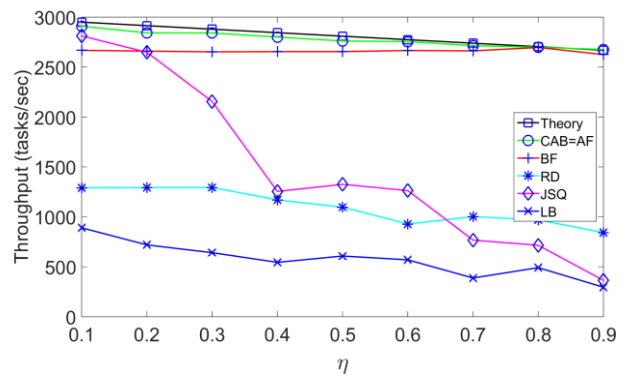

**Figure 15: Experimental throughput of all five polices in P2-biased case. Theoretical optimal (CAB) throughput is also plotted. CAB=AF is indeed optimal and very close to the theoretical value.**

### 7.4 General-symmetric

According to the theoretical results, when the system is general-symmetric, CAB will choose BF policy. Therefore, we pick quicksort-500 and NN-2000. Again, we launch N = 20 benchmarks and sweep over the nine $\eta$ values.

Figure 16 shows the experimental throughput of all the policies as well as the theoretical optimal (CAB) throughput. CAB again outperforms the others, e.g., LB by 2.37x to 4.48x, and the experimental and theoretical results match.

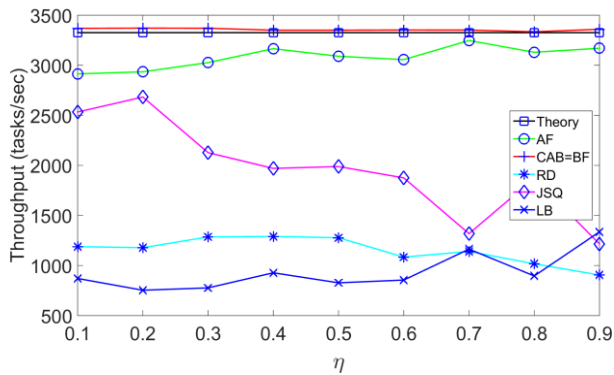

**Figure 16: Experimental throughput of all five polices in general-symmetric case. Theoretically optimal (CAB) throughput is also plotted. CAB=BF is indeed optimal here and is very close to the theoretical value.**

## 8. CONCLUSION AND FUTURE WORK

In this paper, we mathematically formulate the optimal task scheduling policy for heterogeneous systems as an integer non-linear optimization problem with linear constraints, by using queueing theory. We analytically solve for the CAB policy for systems with two processor types, e.g., CPU+GPU. CAB is surprisingly simple, but very general and practical. It covers a wide range of system configurations and is independent of the task size distribution and processor's processing order. Task arrival process is not involved in our modeling, and therefore, no assumption on arrival process is required. For the general case of heterogeneous systems with any number of processor types, we designed the GrIn algorithm which is nearly optimal (within **1.6%** of the optimal), fast, and scalable. GrIn is as general and practical as CAB. Extensive simulations and real platform experiments verify the correctness and generality of our policy. Comparing to classic policies like load balancing, our results, e.g., CAB, can deliver **1.08x** to **2.24x** better performance or **1.08x** to **2.26x** better energy efficiency in simulations, and **2.37x** to **9.07x** better performance in experiments. Future work includes implementing GrIn in a real system with more than two types of processors and deploying our policy in a real operating system kernel.

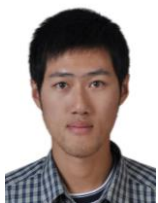
**Zhuo Chen** received the B.S. degree in Electronics Engineering and Computer Science from Peking University, Beijing, China in 2013. He is currently working toward the Ph.D. degree in the Department of Electrical and Computer Engineering, Carnegie Mellon University, Pittsburgh, PA. His research interests are in the area of energy-aware computing. In particular, his research focuses on multi-core heterogeneous/homogeneous system optimization, and low-power application-specific system design.

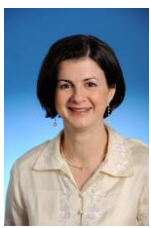
**Diana Marculescu** is a Professor of Electrical and Computer Engineering at Carnegie Mellon University. She has won several best paper awards in top conferences and journals. Her research interests include energy-, reliability-, and variability-aware computing and CAD for non-silicon applications. She is an IEEE fellow.